\documentclass[conference,10pt]{IEEEtran}
\usepackage[ruled,vlined]{algorithm2e}
\usepackage{amsmath}
\usepackage{amssymb}
\usepackage{cite}
\usepackage{color}
\usepackage{float}
\usepackage[T1]{fontenc}
\usepackage{graphics}
\usepackage{epsfig}
\usepackage{microtype}
\usepackage[caption=false]{subfig}
\usepackage{textcomp}
\restylefloat{table}
\usepackage{setspace}

\begin{document}
\title{Distributed Proximal Policy Optimization for Contention-Based Spectrum Access}
\author{
\IEEEauthorblockN{Akash Doshi and  Jeffrey G. Andrews}
\IEEEauthorblockA{Department of Electrical and Computer Engineering, University of Texas at Austin, TX 78712}}
\maketitle 
\normalsize
\begin{abstract}
The increasing number of wireless devices operating in unlicensed spectrum motivates the development of intelligent adaptive approaches to spectrum access that go beyond traditional carrier sensing. We develop a novel distributed implementation of a policy gradient method known as Proximal Policy Optimization modelled on a two stage Markov decision process that enables such an intelligent approach, and still achieves decentralized contention-based medium access. In each time slot, a base station (BS) uses information from spectrum sensing and reception quality to autonomously decide whether or not to transmit on a given resource, with the goal of maximizing proportional fairness network-wide. Empirically, we find the proportional fairness reward accumulated by the policy gradient approach to be significantly higher than even a genie-aided adaptive energy detection threshold. This is further validated by the improved sum and maximum user throughputs achieved by our approach. 
\end{abstract}

\begin{IEEEkeywords}
Medium access, proximal policy optimization, contention, deep reinforcement learning
\end{IEEEkeywords}

\maketitle
\section{Introduction}
Spectrum sharing attempts to allow different transmitters to operate on the same allocated resource (spectrum/time) in a fair manner, while also providing high throughput. To alleviate spectrum constraints, 3GPP \cite{3gpp.36.889} standardized License Assisted Access (LAA) for LTE, and more recently released a study on 5G New Radio Unlicensed (NR-U) \cite{3gpp.38.889}. The approach adopted to access unlicensed spectrum in LAA/NR-U is known as Listen-Before-Talk (LBT) \cite{3gpp.36.889}\cite{etsi} and requires each transmitter to perform a Clear Channel Assessment (CCA) before accessing spectrum i.e. a BS is allowed to transmit on a channel only if the energy level in the channel is less than the CCA threshold level for the duration of the CCA observation time \cite{etsi}. However, a CCA threshold level based MAC decision -- also referred to as an energy detect (ED) threshold in 5G NR -- does not actually reflect the  quality of reception (SINR) at the UE. 

An optional collision reduction scheme known as Request-
to-Send/Clear-to-Send (RTS/CTS) is supported by 802.11, but is known to inhibit potentially successful transmissions, and introduce significant additional overhead and latency \cite{sobrinho2005rts}.  Nearly all WiFi systems disable RTS/CTS.  Recently, multi-agent reinforcement learning (RL) has been applied to design state-based policies that can improve the performance of unlicensed spectrum sharing \cite{lunden2011reinforcement,naparstek2018deep,tonnemacher2019machine}. Most recently, \cite{naderializadeh2021resource} presented a robust and scalable distributed RL design for radio resource management to mitigate interference. None of these papers thus far have attempted to model the asynchronous nature of the decisions made by the transmitters owing to contention. In \cite{Dosh2101:Deep}, we developed a distributed deep RL spectrum sharing algorithm incorporating contention-based medium access. It deployed Deep Q Networks (DQN) at each BS that sequentially decide whether or not to transmit, with the goal of maximizing proportional fairness (PF) network-wide. However, it suffered from slow training convergence and stability, and achieved a much smaller reward than a PF-based BS scheduler.

Policy gradient methods \cite{sutton2018reinforcement} are known to achieve significantly faster convergence than DQN algorithms, while also improving the reward earned by agents in a multi-agent environment. Consequently, in this paper, we design a novel distributed version of a recent policy gradient method known as Proximal Policy Optimization \cite{schulman2017proximal} to optimize medium access under the constraint of a contention-based access mechanism. We employ the paradigm of \textit{centralized learning} with \textit{decentralized execution}, such that each BS will decide whether and how to transmit based only on its own observations.

\section{Problem Statement and System Model} \label{sec:PF_scheduler}
\begin{figure}
    \centering
    \includegraphics[width = 3.4in,height=1.4in]{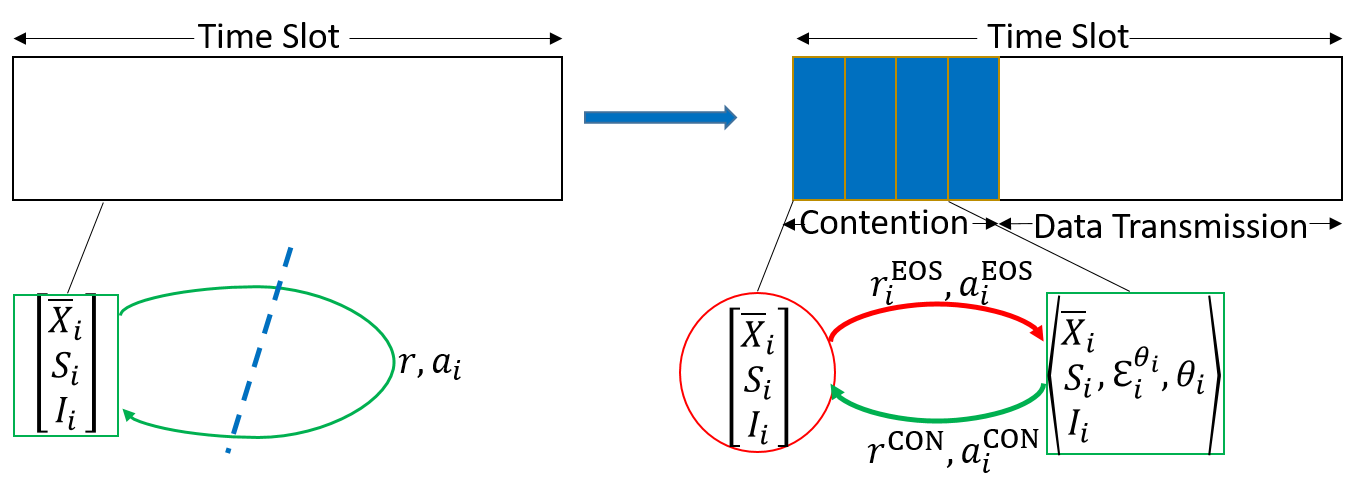}
    \caption{The 2 state MDP at each agent capturing the actions taken and reward obtained on transitioning between the End-Of-Slot(EOS) and Contention(CON) states}
    \label{fig:MDP}
\end{figure}

We consider a downlink cellular deployment of $N$ BSs, with a single UE scheduled per time slot per BS. The notation utilized henceforth is summarized in Table \ref{tab:decpomdp_notation}. Assuming that the UE throughput $R_j[n]$ in each time slot $n$ is approximated by the Shannon capacity $W\log_2(1+\mathrm{SINR}_j[n])$, the same UE is scheduled for reception for $L$ consecutive time slots and each BS transmits at a constant power, the MAC algorithm at each BS has to decide whether or not to transmit to the UE in each time slot. We consider a simplified contention-based access mechanism in which each time slot is divided into a contention and data transmission period. At the start of the contention period consisting of $N$ mini-slots, BS $i$ draws a random counter $\theta_i \in \{0, \ldots ,N-1\}$, with the possibility that $\theta_i = \theta_j$ for $i \neq j$. The counter is decremented by 1 every mini-slot and when this counter expires, the BS ascertains if the channel is clear before transmitting a unique preamble for the remainder of the contention period, followed by data transmission, with the objective of each BS being to maximize the long-term data throughput seen by the UE. Mathematically, in \cite{wang2010scheduling}, this is proved to be equivalent to
\begin{align}\label{eq:PF_problem}
    &\max_{n\rightarrow \infty} \sum_{j=1}^{N}\log(\overline{X}_j[n]) \\
    \mathrm{where} ~~ \overline{X}_j[n] = &~ (1-1/B)\overline{X}_j[n-1] + (1/B)R_j[n].
\end{align}
While Proportional Fair (PF) scheduling would amount to an iterative BS scheduler computing the rate vector $\mathbf{R}^*[n]$ for every time slot $n$, such that
\begin{equation} \label{eq:PF_scheduler}
    \mathbf{R}^*[n] = \underset{\mathbf{R}[n]}{\mathrm{arg\ max\ }} \sum_{j=1}^{N} \frac{R_j[n]}{\overline{X}_j[n]},
\end{equation}
it requires a centralized controller and hence is not realizable in any practical decentralized deployment. In \cite{Dosh2101:Deep}, we formulated \eqref{eq:PF_problem} as a decentralized partially observable Markov decision process - \textit{DEC-POMDP}- with the observation $o_i$ received by BS $i$ given by $o_i[n] = \langle \overline{X}_i[n-1],S_i[n-1],I_i[n-1] \rangle$
and a novel per-timestep reward structure given by $r[n] = \sum_{j=1}^{N} r_j[n] \ \forall \ n > 0$ and $r[0] = \sum_{j=1}^{N} \log(\overline{X}_j[0])$ where 
\begin{equation} \label{eq:per_ts_rwd}
    r_j[n] = \log\Bigg((1-1/B)\Big(1 + \frac{R_j[n]}{(B-1)\overline{X}_j[n-1]}\Big)\Bigg).
\end{equation}
We then incorporated contention by partitioning the 1-state MDP into 2 states, End-Of-Slot (EOS) and Contention (CON), as shown on the right in Fig. \ref{fig:MDP}, with $\mathbf{o}^{\mathrm{EOS}}_i = o_i$ and $\mathbf{o}^{\mathrm{CON}}_i = \langle o_i, \mathcal{E}_i^{\theta_i}, \theta_i \rangle$, where $\mathcal{E}_i^{\theta_i} = \{\mathcal{E}_{ij}^{\theta_{i}}\}_{j\in[N]}$, such that $\mathcal{E}_{ij}^{\theta_{i}}$ is the energy measured at BS $i$ due to an ongoing transmission between BS $j$ and UE $j$. We have $r^{\mathrm{CON}}[n] = r[n]$ and $a_i^{\mathrm{CON}} = a_i$, while both $r^{\mathrm{EOS}}_i$ and $a^{\mathrm{EOS}}_i$ default to 0. The $\mathcal{E}_i^{\theta_i}$ vector served as a message exchanged between agents, and along with the addition of an LSTM layer to each neural network in the system, helped to overcome partial observability in a multi-agent environment \cite{hausknecht2015deep} \cite{foerster2016learning}. 
\begin{table}
    \footnotesize
    \caption[Notation] {UE $j$ and BS $i$ Notation}
	\label{tab:decpomdp_notation}
	\begin{tabular}{ |p{0.6cm}|p{2.5cm}|}
	    \hline
	    $S_j$ & Signal Power\\\hline
	    $I_j$ & Interference Power\\\hline
	    $g_{ij}$ & Path gain with BS $i$ \\\hline
	    $R_j$ & Data Rate \\\hline
	    $\overline{X}_j$ & Average Rate\\\hline
	\end{tabular}
	\hspace{0.2in}
	\begin{tabular}{ |p{0.4cm}|p{3cm}|}
		\hline
		$a_i$ & Action chosen $= \{0,1\}$\\\hline
		$o_i$ & Local observation  \\\hline
	    $g'_{ij}$ & Path gain with BS $j$ \\\hline
		$\pi_i$ & Policy to choose $a_i$ \\\hline
		$W$ & Channel Bandwidth \\\hline
	\end{tabular}
\end{table}

\section{Proximal Policy Optimization (PPO) for Medium Access DEC-POMDP}
\subsection{Proximal Policy Optimization}
Given a state $s$ and an action $a$, we have three terms associated with a typical single agent RL problem: $\pi(a|s;\Theta)$, $\mathcal{Q}^{\pi}(s,a)$ and $V^{\pi}(s)$. We denote by $\pi(a|s;\Theta)$ a policy parameterized by $\Theta$ that returns the probability of an agent selecting action $a$ in state $s$. If an agent starts from state $s$, chooses action $a$ and thereafter follows $\pi$, the expected reward accumulated is represented by $\mathcal{Q}^{\pi}(s,a)$. Finally, $V^{\pi}(s)$ denotes the expected reward accumulated by an agent following $\pi$ starting from state $s$. Policy-based model-free methods directly parameterize the policy $\pi(a|s;\Theta)$ and update $\Theta$ by performing gradient ascent on $J(\Theta) = \mathbb{E}[\gamma^n r[n]]$. The gradient is given by $\nabla_{\Theta} \log \pi(a|s;\Theta) \mathcal{Q}^{\pi}(a|s)$. To reduce the variance of this unbiased estimate of $\nabla_{\Theta} J(\Theta)$, a learnt baseline $b(s) \approx V^{\pi}(s)$ is subtracted \cite{williams1992simple} such that
\begin{equation}
    \nabla_{\Theta} J(\Theta) = \nabla_{\Theta} \log \pi(a|s;\Theta) (\mathcal{Q}^{\pi}(a|s) - V^{\pi}(s)),
\end{equation}
where $A(s,a) = \mathcal{Q}^{\pi}(a|s) - V^{\pi}(s)$ is known as the \textit{advantage}. This approach can then be viewed as an actor-critic architecture, where actor refers to $\pi(a|s)$ while $V^{\pi}(s)$ is the critic. In actor-critic algorithms, both the actor and critic are represented by two separate neural networks, $\pi(a|s;\Theta)$ and $V(s;\vartheta)$,  parameterized by $\Theta$ and $\vartheta$ respectively, and the following loss function is minimized
\begin{equation} \label{eq:loss_ac}
    L(\Theta,\vartheta) = - J(\Theta) + (V(s,\vartheta) - V^{\mathrm{target}}(s))^2.
\end{equation}
A truncated version of generalized advantage estimation (GAE) \cite{schulman2015high} is traditionally utilized to compute $V^{\mathrm{target}}(s_t)$ as
\begin{align}
    & V(s_t;\vartheta) + \delta_t + (\gamma\lambda) \delta_{t+1} + \ldots + (\gamma\lambda)^{T-t+1} \delta_{T-1}, \label{eq:gae}\\
    & \mathrm{where} ~~~ \delta_t = r_t + \gamma V(s_{t+1};\vartheta) - V(s_t;\vartheta). \label{eq:delta_eq}
\end{align}

Proximal Policy Optimization (PPO) \cite{schulman2017proximal} alters the loss function in \eqref{eq:loss_ac} in two ways. Firstly, it replaces $J(\Theta) = \mathbb{E}[\pi(a|s;\Theta) A(s,a)]$ with 
\begin{align}
    &L^{\mathrm{CLIP}}(\Theta) = \mathbb{E}[\min(r(\Theta) A, \mathrm{clip}(r, 1 - \epsilon, 1 + \epsilon) A)], \label{eq:l_clip}\\
    &\mathrm{where} ~~~ r(\Theta) = \pi(a|s;\Theta) /  \pi(a|s;\Theta_{\mathrm{old}}). \label{eq:policy_ratio}
\end{align}
The motivation for this modified metric is to not reward excessively large policy updates, which is enforced via the clip function, while the ``surrogate objective" $r(\Theta)A$ arises from a policy gradient approach known as trust region policy optimization (TRPO) \cite{schulman2015trust}, a precursor to PPO. The term $\pi(a|s;\Theta_{\mathrm{old}})$ in \eqref{eq:policy_ratio} is a constant that corresponds to the evaluation of the policy $\pi$ at the given $(s,a)$ using the current weights $\Theta_{\mathrm{old}}$ of the policy NN. Secondly, PPO adds an entropy bonus $S[\pi(.|s;\Theta)]$ to ensure sufficient exploration. Consequently, the final PPO objective $L^{\mathrm{PPO}}(\Theta,\vartheta)$ which is maximized each iteration is given by \cite{schulman2017proximal}
\begin{equation} \label{eq:PPO}
     L^{\mathrm{CLIP}}(\Theta) - c_1(V(s,\vartheta) - V^{\mathrm{target}}(s))^2 + c_2S[\pi(.|s;\Theta)].
\end{equation}

\subsection{Adapting PPO to a Medium Access DEC-POMDP} \label{subsec:ppo_ma}
We now define $N$ loss functions $L(\Theta_i^{\pi_{\mathrm{CON}}},\vartheta_i^{V_{\mathrm{CON}}},\vartheta_i^{V_{\mathrm{EOS}}})$ corresponding to each BS $i$, by adapting \eqref{eq:PPO} to the 2-state MDP presented in Fig. \ref{fig:MDP}, as follows
\begin{align} \label{eq:loss_ss_ppo}
     L^{\mathrm{PPO}}(\Theta_i^{\pi_{\mathrm{CON}}},\vartheta_i^{V_{\mathrm{CON}}}) - c_{3}L^{\mathrm{VF}}(\vartheta_i^{V_{\mathrm{EOS}}}), 
\end{align}
at each BS $i$, where 
\begin{align}
L^{\mathrm{VF}}(&\vartheta_i^{V_{\mathrm{EOS}}}) = (V_i^{\mathrm{EOS}}(\mathbf{o}_i^{\mathrm{EOS}}) - V_i^{\mathrm{target,EOS}}(\mathbf{o}_i^{\mathrm{EOS}}))^2,
\end{align}
while a similar expression for $L^{\mathrm{VF}}(\vartheta_i^{V_{\mathrm{CON}}})$ is already part of $L^{\mathrm{PPO}}(\Theta_i^{\pi_{\mathrm{CON}}},\vartheta_i^{V_{\mathrm{CON}}})$. Note that the only key changes from \eqref{eq:PPO} are that we have added a term for training $V_i^{\mathrm{EOS}}$ and the input to the EOS and CON NNs will be $\mathbf{o}^{\mathrm{EOS}}_i$ and $\mathbf{o}^{\mathrm{CON}}_i$ respectively, instead of the full system state $s$. To overcome this partial observability, an LSTM layer is introduced in $V_i^{\mathrm{CON}}$, $V_i^{\mathrm{EOS}}$ and $\pi_i^{\mathrm{CON}}$ at every BS $i$. Now, to compute $V_i^{\mathrm{target,CON}}$ and $V_i^{\mathrm{target,EOS}}$, we first observe that \eqref{eq:delta_eq}, when applied to the EOS-CON transition yields
\begin{align} 
    \delta_{i,n}^{\mathrm{CON}} &= r[n] + \gamma^{\frac{1}{2}}V_i^{\mathrm{EOS}}(\mathbf{o}_i^{\mathrm{EOS}}[n+1]) - V_i^{\mathrm{CON}}(\mathbf{o}_i^{\mathrm{CON}}[n]) \label{eq:delta_CON}\\
    \delta_{i,n}^{\mathrm{EOS}} &= \gamma^{\frac{1}{2}}V_i^{\mathrm{CON}}(\mathbf{o}_i^{\mathrm{CON}}[n]) - V_i^{\mathrm{EOS}}(\mathbf{o}_i^{\mathrm{EOS}}[n]). \label{eq:delta_EOS}
\end{align}
Note that the factor of $\gamma^{\frac{1}{2}}$ in \eqref{eq:delta_CON} and \eqref{eq:delta_EOS} is simply meant to keep the overall discount factor to $\gamma$ in one time step.
Substituting \eqref{eq:delta_CON} and \eqref{eq:delta_EOS} into \eqref{eq:gae} and replacing $T$ by the episode length $L$, we obtain
\begin{multline}
     V_i^{\mathrm{target,EOS}} = V_i^{\mathrm{EOS}}(\mathbf{o}_i^{\mathrm{EOS}}) + \delta_{i,n}^{\mathrm{EOS}} + (\gamma^{\frac{1}{2}}\lambda)\delta_{i,n}^{\mathrm{CON}} \\+ \ldots + (\gamma^{\frac{1}{2}}\lambda)^{L-n+1}\delta_{i,L-1}^{\mathrm{EOS}}  \label{eq:V_EOS}
\end{multline}
\begin{multline}
     V_i^{\mathrm{target,CON}} = V_i^{\mathrm{CON}}(\mathbf{o}_i^{\mathrm{CON}}) + \delta_{i,n}^{\mathrm{CON}} + (\gamma^{\frac{1}{2}}\lambda)\delta_{i,n+1}^{\mathrm{EOS}} \\+ \ldots + (\gamma^{\frac{1}{2}}\lambda)^{L-n+1}\delta_{i,L-1}^{\mathrm{CON}}. \label{eq:V_CON}
\end{multline}
Note that we will denote $V_i^{\mathrm{target,CON}}(\mathbf{o}_i^{\mathrm{CON}}) - V_i^{\mathrm{CON}}(\mathbf{o}_i^{\mathrm{CON}})$ in \eqref{eq:V_CON} as $\hat{A}_i^{\mathrm{CON}}$, an estimate of $A_i^{\mathrm{CON}}$. This will be utilized for computing $L^{\mathrm{CLIP}}(\Theta_i^{\pi_{\mathrm{CON}}})$ via \eqref{eq:l_clip}.

\subsection{Generating an episode} \label{subsec:gen_episode}
An episode refers to a collection of $L$ consecutive time slots. At the beginning of time slot $n$, a random counter $\theta_i$ is drawn for each BS $i$. Each  $\pi^{\mathrm{CON}}_i$ outputs two probabilities corresponding to the actions 0 and 1, with 
\begin{equation} \label{eq:action_policy}
    a_i = \underset{a \in A_i} {\mathrm{arg\ max\ }} \pi^{\mathrm{CON}}_i (\mathbf{o}^{\mathrm{CON}}_i)[a].
\end{equation}
While generating an episode during the training of the algorithm, we simply sample the action randomly from the probability distribution outputted by $\pi_i^{\mathrm{CON}}(\mathbf{o}_i^{\mathrm{CON}})$ \cite{mnih2015human}. 

Consider as an example $N = 3$ with BS 0, 1 and 2 being allocated counter values $\langle \theta_0, \theta_1, \theta_2 \rangle = \langle 2, 0, 1 \rangle$ in time slot $n$. Since $\theta_1 = 0$, BS 1 goes first and measures the energy from ongoing transmissions to compute $\mathcal{E}_1^{\theta_1}$. It senses no other BS's transmitting ($\mathcal{E}^{\theta_1}_{1} = [0,0,0]$), and in combination with $\overline{X}_1[n-1]$, $S_1[n-1]$ and $I_1[n-1]$ of the UE it serves, it determines $a_1[n]$ using the policy given in \eqref{eq:action_policy}. Let us assume it chose to transmit (transmission is not a given simply because $\mathcal{E}^{\theta_1}_{1} = [0,0,0]$). BS 2 is scheduled next, detects BS 1 is transmitting such that $\mathcal{E}^{\theta_2}_{21}$ is non-zero and $\pi^{\mathrm{CON}}_2$ instructs it not to transmit. Finally BS 0 also detects a non-zero $\mathcal{E}^{\theta_0}_{01}$, but chooses to transmit. Note that while the training procedure, elaborated in Section \ref{sec:sim_details}, will require training $V^{\mathrm{CON}}_i$, $V^{\mathrm{EOS}}_i$ and $\pi^{\mathrm{CON}}_i$, testing the learnt policy using \eqref{eq:action_policy} only requires $\pi^{\mathrm{CON}}_i$.

Once all the BS's have taken an action $a_i$, the action vector $\mathbf{a}$ ($\langle 1,1,0 \rangle$ in this example) and $\{g_{ij}\}$ are used to calculate the reward $r[n]$ and the updated average rates $\mathbf{\overline{X}}[n]$. These determine the observations $\mathbf{o}^{\mathrm{EOS}}_i$ for the next time slot. In the next time slot $n+1$, a new counter $\theta'_i$ is drawn at each BS $i$ and the process repeated.

\section{Simulation Details} \label{sec:sim_details}
\begin{table*}
    \footnotesize
	\caption[Simulation Parameters] {Simulation Parameters}
	\label{tab:sim_param}
	\centering
	\subfloat[Data Generation Parameters]{\begin{tabular}{ |p{3.5cm}|p{2cm}|}
		\hline
		$N$ & 4\\\hline
		Layout & InH-Office \cite{3gpp.38.889} \\\hline
		Noise PSD & -174 dBm/Hz \\\hline
		Bandwidth $W$ & 20 MHz \\\hline
		(UE, BS) Noise Figure & (9, 5) dB\\\hline
		Fading Coefficient $\alpha$ & 0.1\\\hline
		Smoothing Window $B$ & 10\\\hline
		Center frequency $f_c$ & 6 GHz \\\hline
	\end{tabular}\label{tab:dat_gen_param}}
	\hspace{0.2in}
	\subfloat[RL Training Parameters]{
	\begin{tabular}{ |p{4.2cm}|p{3.8cm}|}
		\hline
		Initial Learning Rate $\eta$ & L1, L2, L3: $(4,4,2) \times10^{-4}$ \\\hline
		Learning Rate Decay &  L1 \& L2: 0.85 ~/~ 500 updates ~~~~L3: 0.5 ~/~ 250 updates\\\hline
		Optimizer & Adam\cite{kingma2014adam}\\\hline
		$N_{\mathrm{batch}}$ & 8\\\hline
		Training Iterations & 800\\\hline
		$\epsilon$, $\gamma$, $L$ &  0.2, 1 - 1e-6, 2000\\\hline
		$|state\_h_{i,n}|$ for (PPO, DQN) & (128, 256) \\\hline
	\end{tabular}\label{tab:dqn_training_param}}
\end{table*}
The performance metric is the expected cumulative reward $\sum_{n=0}^{L} \gamma^n r[n]$, with $r[n]$ given by \eqref{eq:per_ts_rwd}. Note that for $\gamma \rightarrow 1$, $\sum_{n=0}^{L} \gamma^n r[n] \rightarrow \sum_{j=1}^{N}\log(\overline{X}_j[L])$. In each iteration, $N_{\mathrm{batch}}$ episodes are generated by $N_{\mathrm{batch}}$ $\pi_i^{\mathrm{CON}}$ (actors) at each BS $i$ acting in parallel. An overview of the training procedure is presented in Algorithm \ref{algo:ppo}, while the simulation parameters are summarized in Table \ref{tab:sim_param}.

Policy gradient methods in multi-agent environments typically exhibit very high variance and perform poorly in absence of both stationarity and the Markov property. Consequently, to stabilize the training and improve the learnt policy, we incorporate a \textit{decentralized actor centralized critic} approach, first proposed in \cite{NIPS2017_68a97503}. The motivation behind this approach is to use extra information to ease training, so long as this information is not used at test time i.e. centralized training with decentralized execution. In Algorithm \ref{algo:ppo}, we observe that only $\pi_i^{\mathrm{CON}}$ is required for generating an episode i.e. at test time. Hence, we change the input to both $V_i^{\mathrm{EOS}}$ and $V_i^{\mathrm{CON}}$ by replacing $\mathbf{o}_i^{\mathrm{EOS}}$ with $\mathbf{s}^{\mathrm{EOS}}$ at each BS $i$. While we defined $\mathbf{o}_i^{\mathrm{EOS}}[n+1] = \langle \overline{X}_i[n], S_i[n], I_i[n] \rangle$, we have $\mathbf{s}^{\mathrm{EOS}}[n+1] = \langle \overline{\mathbf{X}}[n], \mathbf{S}[n], \mathbf{I}[n] \rangle$ i.e. it will contain the average rate, signal and interference power of all UEs in the previous time slot. Hence the input to $V_i^{\mathrm{CON}}$ will be $\langle \mathbf{s}^{\mathrm{EOS}}, \mathcal{E}_i^{\theta_i},\theta_i \rangle$, while the input to $\pi_i^{\mathrm{CON}}$ remains $\mathbf{o}_i^{\mathrm{CON}}$.

In order to have a fair comparison with the DQN algorithm from \cite{Dosh2101:Deep}, we make two changes in the implementation of distributed DQN. Firstly, we alter the input to $\mathcal{Q}_i^{\mathrm{EOS}}$ to $\mathbf{s}^{\mathrm{EOS}}$ in place of $\mathbf{o}_i^{\mathrm{EOS}}$. Secondly, in each iteration, $N_{\mathrm{batch}}$ episodes are generated using the current $\mathcal{Q}_i^{\mathrm{EOS}}$ and $\mathcal{Q}_i^{\mathrm{CON}}$, in place of the replay memories $D^{\mathrm{EOS}}$ and $D^{\mathrm{CON}}$ utilized in \cite{Dosh2101:Deep} that added one episode generated using the current NNs and removed the oldest episode every iteration. Finally, we will also compare with the PF, ED and Adaptive ED baselines. ED allows a BS to transmit only if $\sum_{j=1}^{N} \mathcal{E}_{ij}^{\theta_{i}} < E_0$. We employ $E_0 = -72 \ \mathrm{dBm}$ \cite{3gpp.36.889}. Adaptive ED finds the ED threshold that maximizes $\sum_{n=0}^{L}\gamma^n r[n]$ for the given configuration of UEs from a set of ED thresholds ranging from -22 to -92 dBm.

\begin{algorithm} \label{algo:ppo}
\setstretch{1}
\SetAlgoNoLine
\For{$\mathrm{iteration}=1,2, \ldots $}{
    \For{$\mathrm{actor}=1,2, \ldots, N_{\mathrm{batch}}$}{
        Generate an episode of $L$ time slots as detailed \\in Section \ref{subsec:gen_episode}. \\
        In each time slot, each BS $i$ chooses to transmit \\with probability $\pi_i^{\mathrm{CON}}[1]$
    }
    \For{$i = 1,2, \ldots, N$}{
    Compute $V_i^{\mathrm{target,EOS}}$,$V_i^{\mathrm{target,CON}}$ and  $\hat{A}_i^{\mathrm{CON}}$ \\using \eqref{eq:V_EOS} and \eqref{eq:V_CON} at each time for all actors.\\
    Perform 1 epoch of gradient ascent with batch \\size $N_{\mathrm{batch}}\times L$ on \eqref{eq:loss_ss_ppo} to update weights of \\$\pi_i^{\mathrm{CON}}$, $V_i^{\mathrm{CON}}$ and $V_i^{\mathrm{EOS}}$. }
}
\caption[caption]{Spectrum Sharing Proximal Policy Optimization}
\end{algorithm} 

\section{Summary of Results} \label{sec:Results}
\begin{figure}[!ht]
    \centering
    \subfloat[Layout 1: $l=20$]{\includegraphics[width=3.1in]{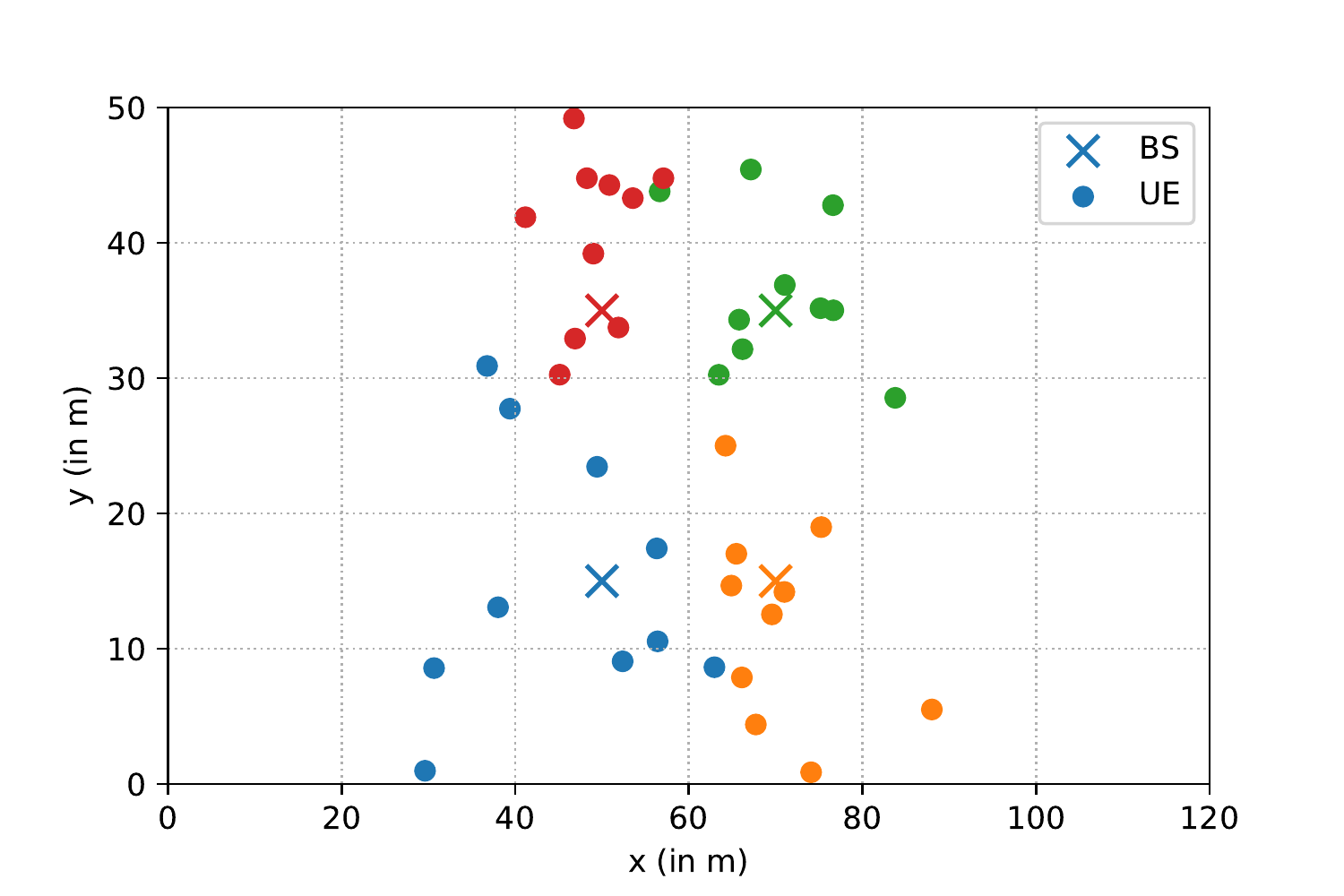}
    \label{fig:layout_1}}
    \hspace{0.2in}
    \subfloat[Layout 2: $l=60$]{\includegraphics[width=3.1in]{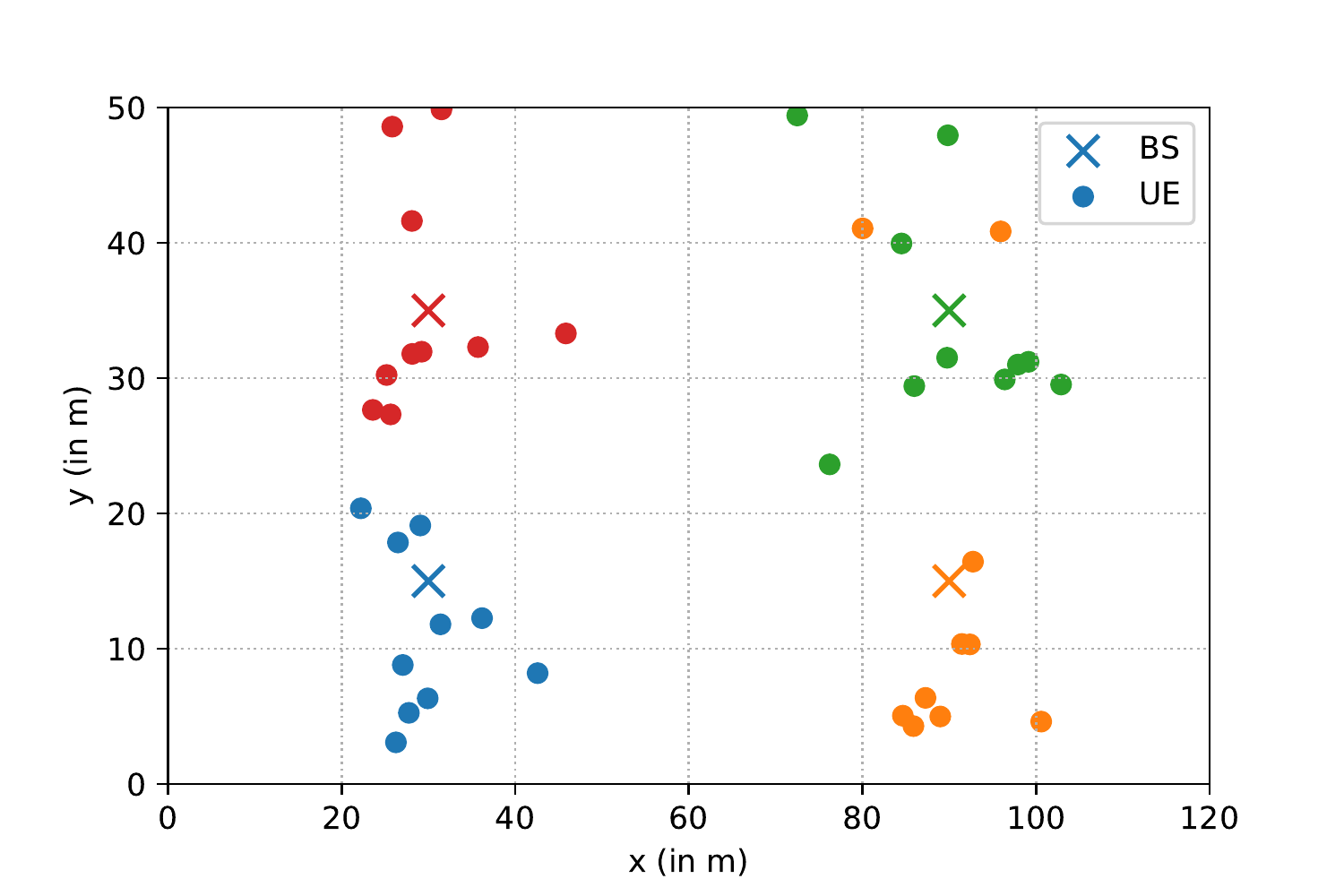}
    \label{fig:layout_3}}
    \caption{Two layouts of 4 BS's at the corners of a $l \times 20$ m rectangle.}
    \label{fig:layout}
\end{figure}

We consider 4 BSs lying at corners of a rectangle of breadth 20 m and length 20 m in \textit{Layout 1} (L1) and 60 m in \textit{Layout 2} (L2). As the rectangle length is increased, for most choice of 4 UE's, the inter-BS energies $\mathcal{E}_i$ will more accurately reflect the quality of reception. This is because the separation between UE's from different BS's reflects the inter-BS separation more accurately as we move from L1 in Fig. \ref{fig:layout_1} to L2 in Fig. \ref{fig:layout_3}.
\begin{figure}[!ht]
    \centering
    \subfloat[Layout 1]{\includegraphics[width=3in]{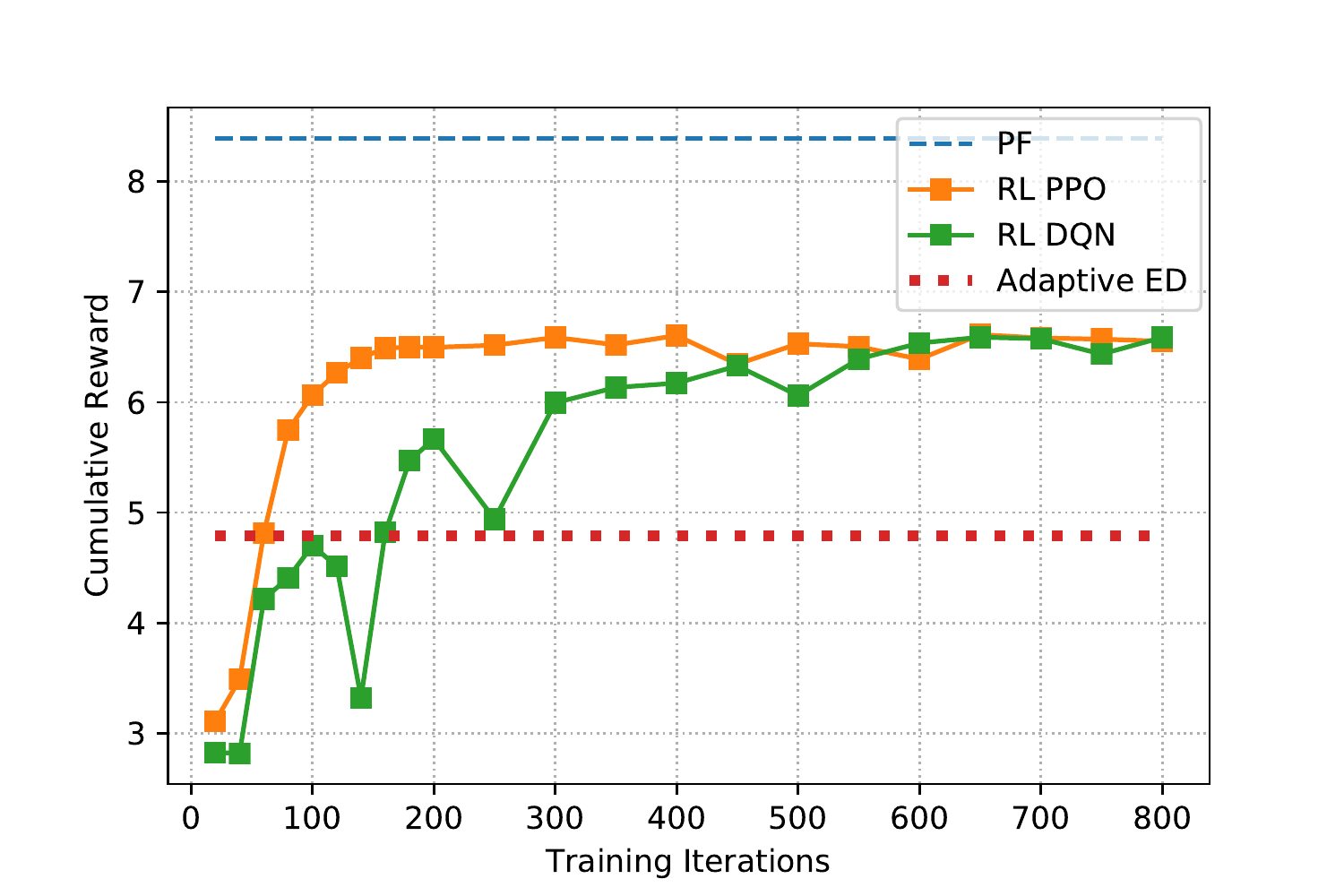}\label{fig:reward_iter_layout1}}
    \hspace{0.2in}
    \subfloat[Layout 2]{\includegraphics[width=3in]{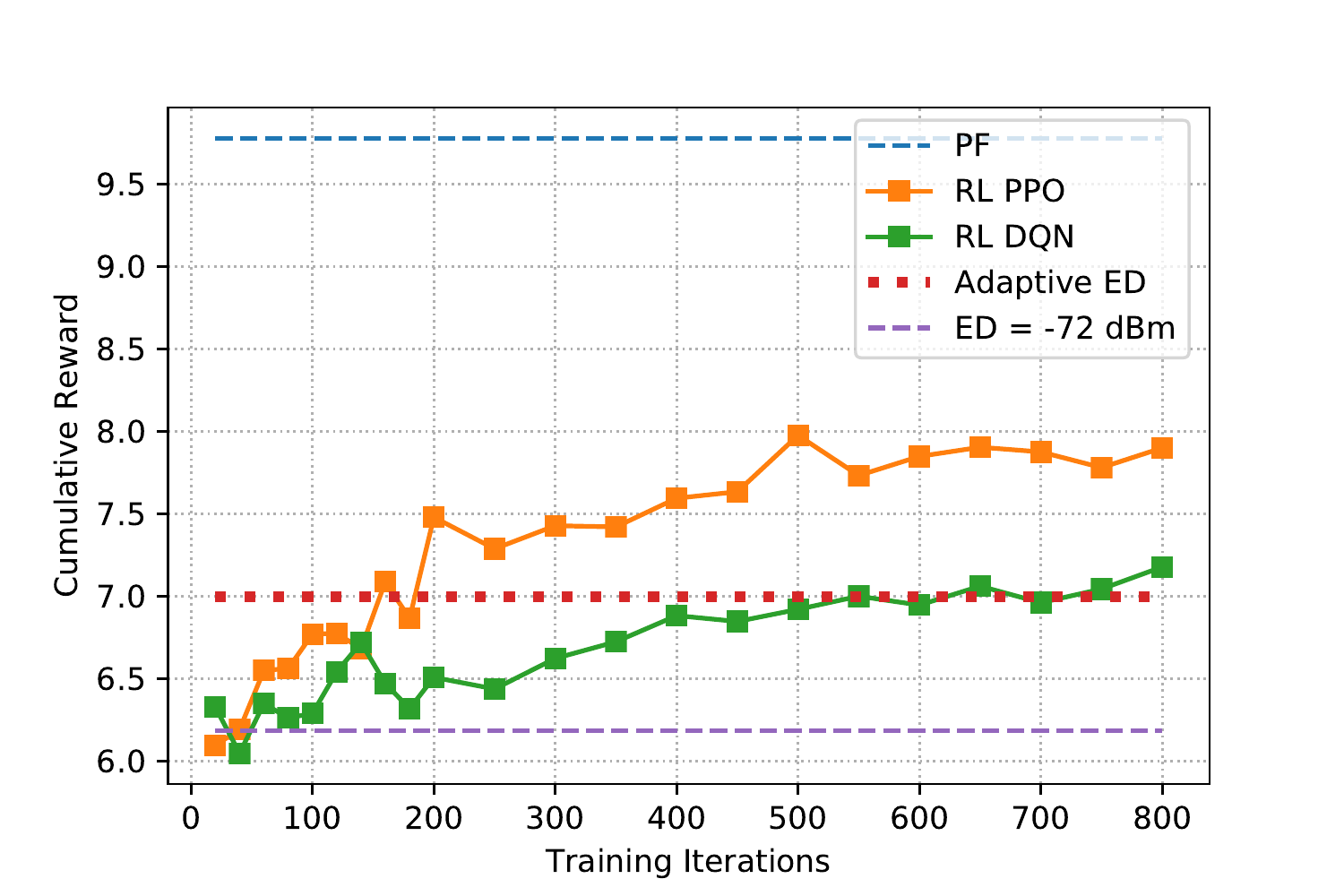}
  \label{fig:reward_iter_layout3}}
    \caption{Cumulative Reward evaluated on Validation Set v/s Training Iterations for \textit{Layout 1} and \textit{Layout 2}}
    \label{fig:reward}
\end{figure}
\begin{figure}
    \centering
    \subfloat[Layout 1]{\includegraphics[width=3in]{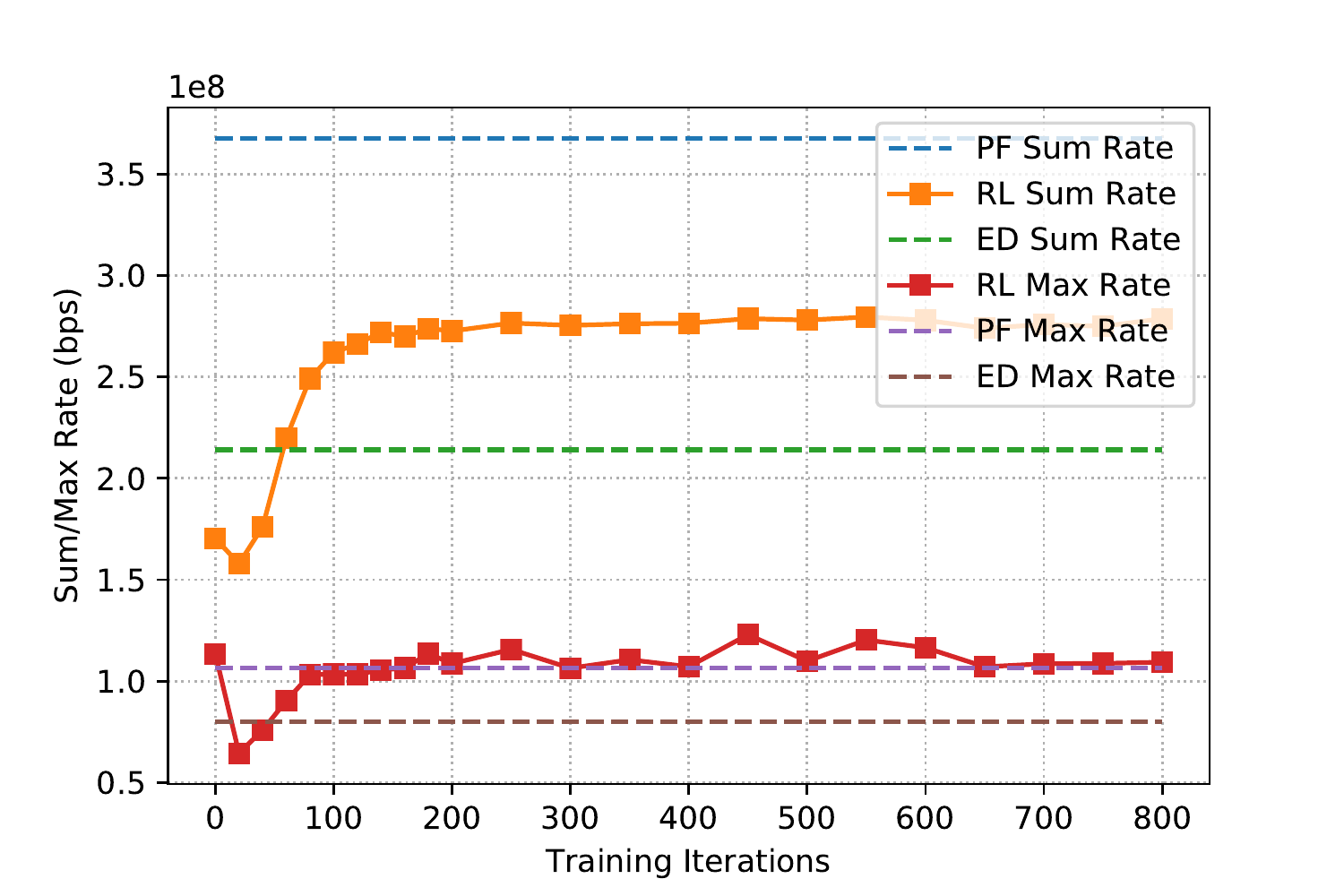}
    \label{fig:reward_layout_1}}
    \hspace{0.2in}
    \subfloat[Layout 2]{\includegraphics[width=3in]{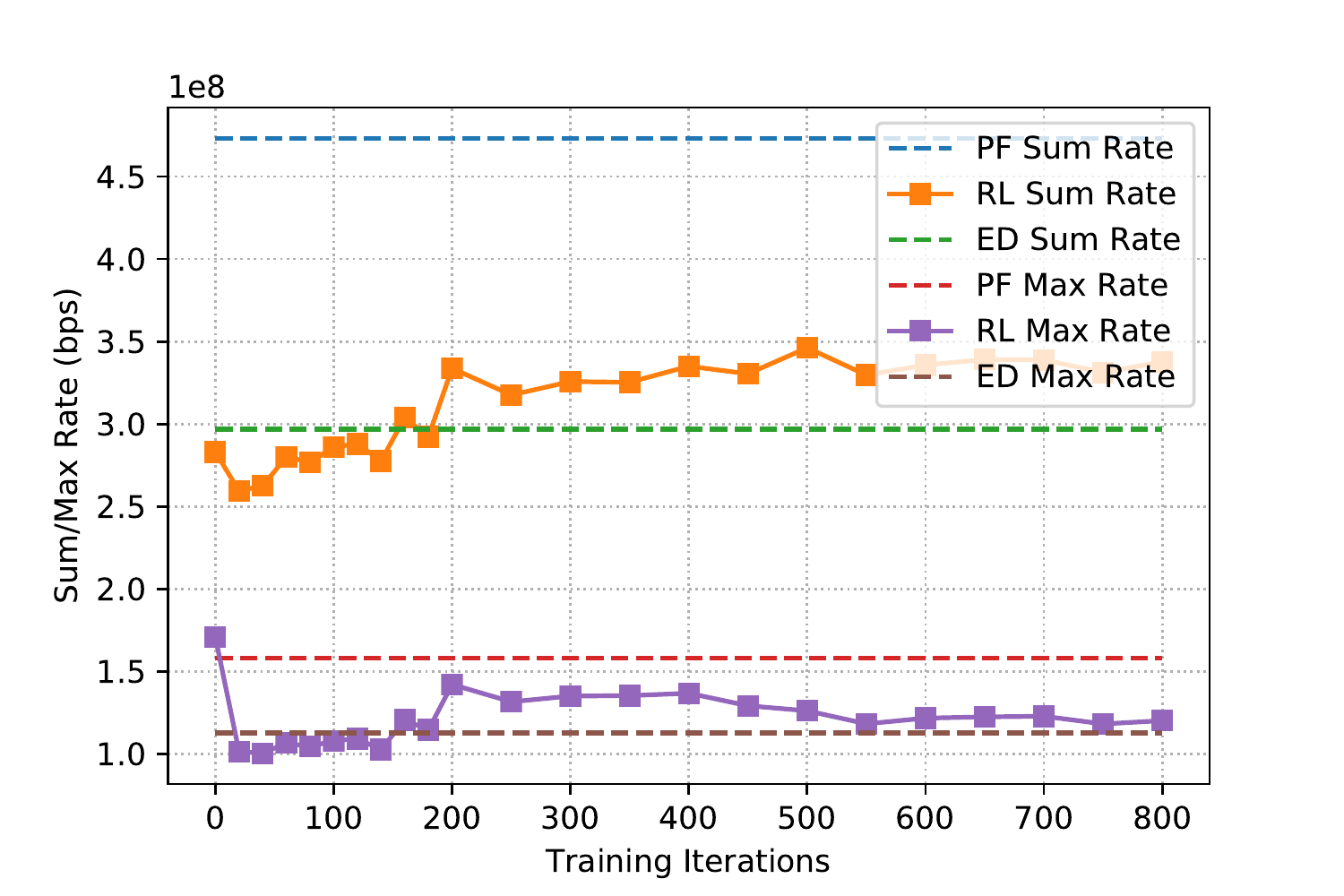}
    \label{fig:reward_layout_3}}
    \caption{Max and Sum UE Rate evaluated on Validation Set v/s Training Iterations for \textit{Layout 1} and \textit{Layout 2}}
    \label{fig:rate}
\end{figure}

The validation curve is shown for Layout 1 and 2 in Fig. \ref{fig:reward_iter_layout1} and \ref{fig:reward_iter_layout3} respectively for both the DQN and PPO methods, along with the constant benchmarks provided by the PF and ED baselines. It is obtained by evaluating the trained models obtained after every 50 iterations on 15 randomly sampled configurations and averaged over 20 realizations of each configuration. Two key observations can be made from the PF and ED baselines: firstly, as we move from L1 to L2, both baselines accumulate a larger cumulative reward. This is because the increasing separation between UEs from different BSs allows more BSs to transmit simultaneously. Secondly, a single standardized threshold of -72 dBm cannot provide the same degree of fairness in different scenarios. In fact, for L1, -72 dBm is a very pessimistic threshold that ends up primarily switching off all the BSs, hence it has not even be plotted. On the other hand, the RL PPO algorithm consistently outperforms even the adaptive ED threshold for all three layouts. More importantly, the RL PPO algorithm tends to always converge faster to the optimal solution than DQN, has a more stable training curve and even significantly outperforms DQN in some instances e.g. Layout 2. 

For every realization of each UE configuration, we compute the sum rate $W \sum_{j=1}^N \overline{X}_j[L]$ and max rate $W \max_{j} \overline{X}_j[L]$ obtained using the RL PPO algorithm at the end of $L$ time-steps. The sum and max rate, averaged over all realizations and configurations, and evaluated using the trained model obtained after every 600 iterations, are plotted for L1 and L2 in Fig. \ref{fig:reward_layout_1} and \ref{fig:reward_layout_3} respectively, along with the corresponding PF and Adaptive ED baselines. Consistent with the higher cumulative rewards earned by the RL algorithm, RL PPO achieves a sum rate at least equal to the adaptive ED algorithm, but always manages to increase the gap between the maximum and sum rate, hence providing a greater degree of fairness than the adaptive ED algorithm.

\section{Conclusions \& Future Directions}
The distributed PPO algorithm designed in this paper jointly utilized the information from LBT-based spectrum sensing at the BS along with the average rate, signal and interference power seen by the UE it serves to determine whether a BS will transmit in the designated time slot. Consequently, it was found to significantly outperform a configuration adaptive ED threshold, and also achieve improved UE throughputs. With a view to the design of a learning based BS, the framework developed in this paper has the potential to be applied in a variety of decision-making problems, including adaptive modulation and coding, beam selection for scheduling, coordinated scheduling and channel selection in the frequency domain. In essence, these extensions tremendously increase the dimensionality of the output action space, from a simple transmit Yes/No decision to a choice of MCS, beamformer, user and subcarrier.

\bibliographystyle{IEEEtran}
\bibliography{bibtex.bib}
\end{document}